\documentstyle[12pt]{article}

\begin{document}
\def\cd{\cdot}
\def\qth{q_{\theta}}
\def\thv{\vec{\theta}}
\def\yv{\vec{y} }
\def\gh{\hat{\gamma} }
\def\gmv{\vec{\gamma} }
\def\alv{\vec{\alpha} }
\def\muv{\vec{\mu} }
\def\muh{\hat{\mu} }
\def\xbb{\overline{\overline{x}} }
\def\ybb{\overline{\overline{y}} }
\def\xb{\overline{x} }
\def\yu{\underline y\,}
\def\Yu{\underline Y\,}
\def\Xb{{\overline{{\bf{X}}}}}
\def\Ytu{\underline \tilde Y\,}
\def\Xtuu{{\bf \tilde X}\,}
\def\PXo{{\bf P}_{X_0}\,}
\def\Omuu{{\bf \Omega}\, }
\def\Siuu{{\bf \Sigma}\, }
\def\SiuuR{{\bf \Sigma}_{RC}\, }
\def\SiuuhR{{ \Siuuhat}_{RC}\, }
\def\Au{\underline A}
\def\alu{\underline\alpha}
\def\thu{\underline{\theta}}
\def\sgb{\overline{\sigma}}
\def\aluhat{ \underline{\hat\alpha} }
\def\beu{\underline\beta}
\def\Auu{{\bf A}\,}
\def\Lauu{{\bf \Lambda}}
\def\yb{\overline{y} }
\def\hb{\overline{h} }
\def\nb{\overline{n} }
\def\Bph{{\hat{B}_p}}
\def\ev{\vec{e}}
\def\xv{\vec{x}}
\def\wv{\vec{w}}
\def\ssgh{{\hat \sigma}^2}
\def\sstg{{\hat \sigma}_{2stg}}
\def\sgstg{{\hat \sigma}_{2stg}}
\def\ssg{\sigma^2}
\def\ybv{\vec{\overline{y}}}
\def\xbv{\vec{\overline{x}}}
\def\xmv{\vec{\overline{x}}}
\def\bvh{\vec{\hat{b}}}
\def\btv{\vec{ \beta}}
\def\btvh{\hat{\vec{ \beta}}}
\def\btd{\vec{ \beta}_{dl}}
\def\bthR{\hat{\vec{ \beta}}_{RC}}
\def\btbv{\vec{\overline{ \beta}}}
\def\btbvh{\hat{\vec{\overline{ \beta}}}}
\def\bvb{\vec{\overline{ \beta}}}
\def\yh{\hat{y}}
\def\xv{\vec{x}}
\def\wv{\vec{w}}
\def\xmv{\vec{\overline{x}}}
\def\bvh{\vec{\hat{ b}}}
\def\bvb{\vec{\overline{ \beta}}}
\def\nl{\hfill\break}       
\def\np{\vfill\eject}       
\def\ns{\vskip 2pc\noindent}       
\def\nsect{\vskip 2pc\noindent}       
\def\nli{\hfill\break\noindent}       
\def\ni{\noindent}
\def\IR{I\kern-.255em R}
\def\Ze{Z_{eff} }
\def\Zeff{Z_{eff} }
\def\tE{\tau_{E} }
\def\nb{\overline{n} }
\def\eps{\epsilon }
\def\llt{\lambda_T }
\def\pbs{\overline{p} }
\def\pbonn{\overline{p_1} }
\def\ptwb{\overline{p_2} }
\def\yu{\underline y\,}
\def\Yu{\underline Y\,}
\def\Xuu{{\bf X}\,}
\def\Siuu{{\bf \Sigma}\, }
\def\Siuuhat{{\bf {\hat\Sigma}}\, }
\def\XSX{{\Xuu^t \Siuuhat^{-1} \Xuu}\, }
\def\Dlh{{\bf {\hat\Delta}}\, }
\def\Suu{{\bf S}\, }
\def\muu{\underline{\mu}\, }
\def\muuhat{\underline{\hat\mu}\, }
\def\muuhathat{\underline{\hat{\hat\mu}}\, }
\def\eu{\underline e\,}
\def\wdr{w_{d,r}}
\def\Wuu{{\bf W}}
\def\Au{\underline A}
\def\Bu{\underline B\,}
\def\Eu{\underline E\,}
\def\alu{\underline\alpha}
\def\aluz{{\underline\alpha}^{(0)}}
\def\aluhat{ \underline{\hat\alpha} }
\def\eps{ \epsilon}
\def\bv{\vec \beta}
\def\beu{\underline\beta}
\def\cur{{\underline c}_r}
\def\curt{{\underline c}_r^t}
\def\euu{{\bf e}\,}
\def\Auu{{\bf A}\,}
\def\Buu{{\bf B}\,}
\def\Cuu{{\bf C}\,}
\def\Euu{{\bf E}\,}
\def\Iuu{{\bf I}\,}
\def\Luu{{\bf L}\,}
\def\Muu{{\bf M}\,}
\def\Puu{{\bf P}\,}
\def\Quu{{\bf Q}\,}
\def\Lauu{{\bf \Lambda}}
\def\Dl{{\bf \Delta}}
\def\Vuu{{\bf V}\,}
\def\Vuuhat{{\bf{\hat V}}\,}
\def\Wd{\dot{W}}

\begin{center}
{\bf RANDOM COEFFICIENT H MODE \\
CONFINEMENT SCALINGS}\\

K.S. Riedel \\
Courant Institute of Mathematical Sciences \\
New York University \\
New York, New York 10012
\end{center}

\begin{abstract}
he random coefficient two-stage regression algorithm with the collisional Maxwell-Vlasov constraint is applied to the ITER H-mode confinement database. The data violate the collisional Maxwell-Vlasov constraint at the 10-30\% significance level, probably owing to radiation losses. The dimensionally constrained scaling, $\tau_E =$ $ 0.07192 M^{1/2}$ $ (R/a)^{-0.221} R^{1.568} \kappaκ^{.3} I_p^{.904} B_t^{.201} \bar{n}^{0.106} P^{-0.493}$, is similar to ITER89P with a slightly stronger size dependence.

\end{abstract}
\np
{\bf I. INTRODUCTION}

The H mode confinement group has assembled an 
excellant H mode global energy confinement database$^1$.
In this letter, we apply the advanced statistical techniques, specifically
the dimensionally constrained
random coefficient (R.C.) regression of Refs. [2-4].

Our analysis differs from the H mode database group in several
significant ways. First, we apply more stringent time stationarity
and reactor relevant selection criteria. We also remove a number
of influential outliers.  The resulting single device scalings are
more uniform and closer to standard L mode confinement scaling.

To model this tokamak to tokamak
variation, we treat the scaling differences between devices as random
variables. 
We begin by estimating a scaling expression
using the random coefficient two step regression procedure of Refs. [2,3].
The collisional Maxwell Vlasov (C.M.V.) similarity is then tested and
imposed$^4$.

The ITER H mode database consists of one large tokamak, JET, one
medium size tokamak, D3D and four small tokamaks, ASDEX, PDX, PBX,
and JFT2-M. Thus the H mode database is significantly more statistically
unbalanced than the corresponding L mode database. 

Since the L mode
database consists of roughly equal numbers of large and small tokamaks,
the L mode size scaling is more accurately determinable. The
H mode dataset includes only one large device, and thus it is also impossible to
assess if the observed confinement dependencies on JET are the
systematic differences from smaller experiments or are purely the
manifestation of tokamak to tokamak differences.
Furthermore, JET has a significantly lower power density and beta
value relative to the Troyon beta limit$^{5}$.

In evaluating the confinement time scaling, we need to decide
which confinement time to use. Most but not all devices provided
both a magnetohydrodynamic (MHD) and a diamagnetic confinement time. The 
H mode confinement group recommended that the MHD confinement
time be used for JFT2-M, D3D, PDX, and PBX-M, that the diamagnetic
confinement time be used for JET and that an MHD equivalent
diamagnetic confinement time be used for ASDEX. 

We follow the
H mode database group's recommendations except that for ASDEX, we use the
geometric average of the diamagnetic confinement time 
and the MHD equivalent confinement time. The MHD confinement
time is notoriously difficult to determine in circular crossection 
devices. In addition, the principal ASDEX group member disapproves 
of the MHD correction advocated by the group as a whole. 
Finally, the MHD confinement time is almost always larger
than the diamagnetic confinement time. By using the average of the
diamagnetic confinement time and the MHD confinement time for ASDEX 
and the MHD confinement time for most other 
machines, we are effectively penalising ASDEX. This small penalty
may compensate in some way for ASDEX's closed divertor. 

In the H mode database, each experimental group calculates their
own thermal confinement time. In our initial analysis, it appears
that the systmatic differences in the evaluation procedures are
sufficiently large to make the determination of a combined thermal
confinement time infeasible. The overall tendency appears  to be to
multiply the total confinement time by $\nb^{.2}$.

We assume that the isotope enhancement factor
is $M^{1/2}$. Unfortunately, no accurate measurement of the
species mixture is available. We adopt the standard convention
that $H \rightarrow D$ discharges have isotope equal $1.5$.
The within tokamak isotope scaling for D3D is roughly $M^{.56}$
and we correct the hydrogen D3D discharges by this amount.
For PDX, the isotope scaling is roughly linear, $M^{1.0}$, which
may be due to poor beam penetration. We therefore exclude all
PDX $H \rightarrow D$ discharges. JFT2-M has virtually no
isotope dependence. Since the parametric dependencies in
JFT2-M are nearly independent of isotope, we force
the JFT2-M discharges to scale as $M^{1/2}$.

\ns
{\bf II. INDIVIDUAL TOKAMAK CONFINEMENT SCALINGS}

In their initial analysis of the H mode database, the H mode confinement
group specified a standard data subset$^1$. Their constraint consist
of limits for the relative radiated power, the relative fast
ion content, the relative time evolution of the plasma energy,
the relative plasma beta and a lower bound on $q_a$. 
We accept all the standard constraints of H mode group. 
We impose the divertor pressure ratio constraint recommended
by PDX. 

In addition,
we impose a number of additional constraints.
Table 1 gives summary data for our data subset.
Throughout this article, we describe the plasma
current , $I_p$,
in units of MAmperes, the toroidal magnetic field, $B_t$ in
units of Teslas, the total heating power $P$ in
MWatts, and the line averaged
plasma density in $10^{19}$ particles per cubic meter.

We restrict our analysis to discharges
with $q_{95} \le 6.5$. 
For ASDEX, we restrict to discharges
with $q_{95} \le 6.0$. 

In restricting the dataset to discharges
where the auxilary heating dominates the Ohmic power, we must determine
the Ohmic power. The H mode database contains $P_{ohm}$ as measured by
the instantaneous loop voltage. However the instantaneous loop voltage
measures the edge toroidal electric field and not the core electric
field. This discrepency can be quite large for nonsteady state plasmas.
Furthermore, we want to restrict the ratio of the Ohmic power before
auxilary heating to the auxilary power. The instantaneous Ohmic power
is relevant for power balance calculations but not for our constraint.
We note that almost all tokamaks observe an Ohmic electric field of
approximately  one Volt over a wide variety of Ohmic conditions. For
the purposes of constraining the data, we define an equivalent Ohmic 
power, $P_{ohm}^* \equiv I_p \times 1 Volt$.

{\it We restrict the relative Ohmic power by $P_{ohm}^*/(P_{abs} + P_{ohm}) < .4$.
Since replacing hot plasma with more cold plasma is usually counterproductive,
we require $\dot{\nb}\tE/\nb < .4$}. We also eliminated half a dozen
JET ELMy discharges at low values of $\nb/I_p$.

The time stationarity  requirement on $\nb$ affects JET most strongly.
The upper limit on $q_a$ affects D3D most strongly.
The requirement on the relative Ohmic power affects JFT2-M and JET more strongly.
Imposing the JET Ohmic power restriction reduces
the very strong $\nb$ and $B_t$ scalings relative to the ``standard" 
JET dataset.


In examining the residuals, $y_i - \yh_i(I_p,B_t,\nb,P)$,
we noticed that in a number of cases the residual errors
depended on the relative time change of the energy. In other
words, the more nonstationary the discharge, the better
the confinement. Therefore we imposed the stronger constraints of
$\Wd/(P_{abs}+P_{Ohm}) <.20$ for PBXM and 
$\Wd/(P_{abs}+P_{Ohm}) <.165$ for PDX. 

The D3D residual errors depend slightly on the normalised
plasma energy change, $\Wd$. More precisely, the D3D discharges
where $\Wd$ is determinable systematically had better 
calculated confinement times then the D3D discharges where
$\Wd$ is indeterminable. In D3D, $\Wd$ is almost always
determinable in elmfree discharges and can seldom be evaluated 
in the ELMy D3D discharges. Thus systematically the 
elmfree D3D discharges received $\Wd$ corrections in the
energy confinement time and the elmy D3D discharges did not.
The elmfree D3D discharges appear to have a slightly better calculated
confinement but this may be an artifact of the $\Wd$ analysis.

Aside from this slight confinement degradation in D3D,
{\it no other tokamak shows any significant evidence
of confinement differences between elmy and elmfree discharges.}
We therefore combine the elmy and elmfree discharges.
    
The $B_t$ dependence of confinement is extremely difficult to 
determine. To zeroth approximation, there is virtually no $B_t$ variation
in JFT2-M, PBX-M or PDX, little $B_t$ variation in ASDEX,
the edge $q$, $q_{95}$
is nearly fixed in D3D and to some extent in JET as well. 
The current and density scalings are coupled in D3D.

Furthermore, a number of influential outliers strongly impact
the $B_t$ scaling in PDX and ASDEX. To examine the $B_t$
dependencies, we regressed the individual tokamaks versus, 
$I_p, \ \nb$ and $P$ and then plotted
the residuals, $y_i - \yh_i(I_p,\nb,P)$, versus $B_t$.

For PDX, the vast majority of the PDX discharges show
only a weak dependence on magnetic field. However several extremely 
low magnetic field discharges suffered from very degradated
confinement.
We assume that these ultra low $B_t$ discharges lie outside the
standard H mode parameter regime and therefore remove these datapoints.

Similarly, ASDEX has one influential datapoint at extremely low
magnetic field which performed extremely well and which was 
determining the ASDEX magnetic field scaling to be $B_t^{-.193}$.
The rest of the ASDEX data indicated a weak to nonexistent 
$B_t$ dependence. Therefore the ultra low $B_t$ points were dropped.

Finally, JFT2-M has a small number (4) of very low $B_t$ discharges 
relative to the mean JFT2-M magnetic field of 1.26 T. This
group of discharges indicates that JFT2-M confinement has a
small positive $B_t$ exponent. Unfortunately, the JFT2-M is
very unbalanced in the $B_t$ covariate with almost all the
data concentrated at 1.26T. Thus no accurate $B_t$ scaling is 
possible for the JFT2-M dataset.  

A comparison of Table 2a with Table 2b (corresponding to
Table X of Ref. 1) shows that our
within tokamak scalings vary significantly less in the
new restricted dataset. 
Thus these restrictions result in a more uniform dataset which better
characterises normal H mode discharges.
Our scalings are almost always closer
to a L mode type scaling than the standard subset of Ref. 1.
We have scrutinised the more ``pathological" scalings more carefully
and therefore have tended to achieve this more uniform, L mode-like
behavior. 

Table 3 summarises how each of our constraints has reduced the number
of discharges.

We agree with the H mode database group that the $B_t$ scaling is
poorly determined as a within tokamak covariate in the present database.
Not only are the $B_t$ scalings sensitive to the outlying datapoints,
but also the root mean square error (RMSE) appears to be an extremely
broad function of the $B_t$ exponent. In cases such as this where the
RMSE is much broader than the half width predicted by ordinary 
least squares (OLS) regression, the at least one of the 
assumptions of OLS regression, such as the correctness of the model
or the independence of the errors, is almost always violated.

Furthermore, determining the standard between tokamak dependencies
of $R, \ R/a, \ \kappa$ and the overall constant is already a 
delicate and questionable procedure for six tokamaks. Treating
$B_t$ as an additional between tokamak covariate is clearly illposed
in the present database.

In our analysis, we exclude the predominately $B_t$ principal component 
in JFT2-M, PDX< and PBX-M. We also exclude the $I_p$ principal
component in PBX-M. We include all the D3D principal components,
however the $I_p, \ B_t$, and $\nb$ scalings are strongly coupled.

\ns

{\bf III. RANDOM COEFFICIENT SCALING}

We begin with an ordinary least square regression analysis of
our 823 datapoint dataset
\footnote{We present our scalings centered about the database
mean, thus the mean values of our database are apparent. Also
if the scaling coefficients are rounded, the overall constant 
in the centered formulation does not need to be adjusted.
The overall constant in the noncentered version should be corrected
to match the overall constant of the centered formulation.}
:

$\tau_E M^{-1/2} =.06371$
$$
\left( {R/a \over 3.804} \right)^{-.217}
\left( {R \over 1.696} \right)^{2.113}
\left( {\kappa \over 1.398} \right)^{.379}
\left( {I_p \over .5667} \right)^{.729} \left(
{B_t \over 1.774} \right)^{.511} \left(
{\nb \over 4.486} \right)^{.090} \left(
{P \over 2.918} \right)^{-.510}
\ .
\eqno (1)$$

The $B_t$ scaling of $B_t^{.511}$ even exceeds the $B_t^{.291}$ 
exponent of JET.
This can only happen when the within tokamak $B_t$
scalings are poorly determined and the root mean squared
error in fitting the corrected mean confinement time of
the tokamaks to the between tokamak covariates, $R,\ R/a,$
and $\kappa$ can be significantly reduced by including
$B_t$ as a basically between tokamak covariate.  

The random coefficient within tokamak scalings are the matrix weighted
average of the scalings of the individual tokamaks. Thus the 
$B_t$ scaling will not exceed the maximum scaling observed in any
tokamak. 

We briefly summarise our random coefficient analysis$^{2-4}$.
First, for each tokamak, a scaling and covariance is estimated in
$I_p, \ B_t, \ \nb $ and $P$. We calculate the empirical mean
and covariance of these within tokamak scalings using the Swamy
random coefficient weighting procedure. Second, the mean confinement
time of each tokamak is corrected for the within tokamak scalings.
The scalings with $R/a,\ \kappa $ and $R$ are estimated by
regressing the corrected mean energy times of the tokamaks.
The error, $\SiuuR$ in our estimate, $\bthR$, of the scaling vector is
given by Eq. (18a) of Ref. 2.

Since three 
of the  tokamaks, JFT2M PDX, and PBX-M
 have virtually no $B_t$ variation, we apply
the projection  missing value algorithm$^4$. 
The projection  missing value algorithm consists of
using only the principal components of the within
tokamak scalings which are estimatable.
Unfortunately, {\it the uncertainty in the $B_t$
scaling direction will be systematically underestimated
since we are unable to compensate for the fewer degrees
of freedom in the $B_t$ direction.} 
In other words, when we estimate the covariance of the $B_t$
exponents and divide through by the number of degrees of freedom,
we use 6 - 1 = 5 instead of 3 - 1 =2.


{\it In the second stage regression, to determine the $R$, $R/a$ and
$\kappa$ scalings, we weight the larger tokamaks, D3D and JET, a
factor of two larger than the smaller tokamaks.} This big tokamak
weighting factor is based on our subjective estimation of the relative
importance bigger tokamaks should have in determining the scaling. 

In the second stage regression, we apply ridge regression
with a relative ridge parameter of $\theta_{nrm}= 0.005$,
a half percent downweighting. This downweighting affects
the $R$ exponent predominately.  

$\tau_E M^{-1/2} =.06298$
$$
\left( {R/a \over 3.804} \right)^{-.171}
\left( {R \over 1.696} \right)^{1.988}
\left( {\kappa \over 1.398} \right)^{.239}
\left( {I_p \over .5667} \right)^{.876} \left(
{B_t \over 1.774} \right)^{.165} \left(
{\nb \over 4.486} \right)^{.075} \left(
{P \over 2.918} \right)^{-.519}
\ .
\eqno (2)$$

The resulting scaling resembles the ITER89P scaling$^6$ except that it
has a very strong size scaling. A principal components analysis of
the random coefficients matrix, $\SiuuR$, reveals that the size scaling
is the most poorly determined exponent. The large variance in the
$R$ exponent is a consequence of the database only containing one
large tokamak. 

The aspect ratio scaling uncertainty is  
relatively small due to the presence of PBX-M.  
The medium size tokamak, D3D, and the large tokamak,
JET, in the database have small aspect ratios. Thus a negative
exponent on the aspect ratio scaling indicates that D3D and JET
have better confinement than a $(R/a)^0$ dependence would indicate.
If a $(R/a)^0$ dependence were required in our scaling, an
even stronger size dependence would result.

The scaling also has  
a noticable component which violates  collisional. 
Maxwell Vlasov similarity. 

\ns
{\bf IV. COLLISIONAL MAXWELL VLASOV CONSTRAINT}

We would like to require that our log linear scaling expression 
be dimensionally consistent with
the  collisional Maxwell Vlasov (C.M.V.) system.
Neglecting the ratio of the Debeye length to all other scale
lengths, the physical
system is prescribed by three dimensionless
variables$^{7}$:
$ \beta \equiv \nb T_i/B_t^2 \ , \rho_i* \equiv (M T_i)^{1/2} / RB_t \ ,
\nu_i* \equiv R \nb q /T_i^2$,
together with the four naturally dimensionless variables: $\kappa, R/a, q_{cyl}$
and $M$. 
As shown in Refs. [2,4,8], the requirement that
$\tE \Omega_i$ is 
a {\it log linear function of the dimensionless variables} can
be treated as a linear constraint on the parameter vector, $\btv$:
$\gmv \cdot \btd + \gamma_B=0$.

Since the random coefficient algorithm not only produces efficient
estimates of the parametric scalings but also a covariance
matrix for the errors in the scaling, we are able to test
this similarity ansatz.

Therefore we determine a C.M.V. constrained scaling within the
multiple tokamak R.C. analysis.
For the $F(1,5)$ distribution, 
the  $50 \% \ $ confidence level (corresponding to the halfwidth)
is  $T^2 = 0.528$,
the  $75 \% \ $ confidence level is  $T^2 = 1.69$,
the  $90 \% \ $ confidence level is  $T^2 = 4.06$.

We find  the test statistic, $T^2 \equiv
| \gmv \cdot \bthR + \gamma_B |^2/ {\gmv^t \cd \SiuuhR \cd  \gmv}
= 1.99$. 
Thus the null hypothesis that the data  can be explained by a 
dimensionless power law scaling can be rejected with slightly more than a $75 \% $
certainty. This conclusion is based on our extremely crude but
selfconsistent modeling of the R.C. covariance. In our L mode
analysis, the corresponding result was $T^2_{Lmode} =0.168$.  
Thus the L mode data not only supported C.M.V. similarity,
but also suggested that the L mode estimate of the dimensional projection,
${\gmv^t \cd \SiuuhR \cd  \gmv}$, was far too large.

The dimensionally constrained H mode scaling is

$\tau_E M^{-1/2} =.06301$
$$
\left( {R/a \over 3.804} \right)^{-.221}
\left( {R \over 1.696} \right)^{1.568}
\left( {\kappa \over 1.398} \right)^{.300}
\left( {I_p \over .5667} \right)^{.904} \left(
{B_t \over 1.774} \right)^{.201} \left(
{\nb \over 4.486} \right)^{.106} \left(
{P \over 2.918} \right)^{-.486}
\ .
\eqno (3)$$

We have determined the dimensionless scaling which is closest to
the unconstrained R.C. scaling {\it measured in the $\Siuu_{RC}^{-1}$
metric}. Since the dominant uncertainty occurs in the $R$ exponent,
our dimensionally constrained scaling differs from the unconstrained
scaling of Eq. 2 by a weaker size scaling. Since $R$ and $R/a$ are
strongly anticorrelated, the aspect ratio scaling decreases as well.

We give the scaling coefficients to three digits accuracy,
not because of precision, but to reduce the extent which rounding
error induces a violation of C.M.V. similarity. The noncentered
version of the constrained scaling law is

$\tau_E = .07192 M^{1/2} 
\left( R/a \right)^{-.221} R^{1.568}
\kappa^{.300} I_p ^{.904} 
B_t^{.201} \nb^{.106} P^{-.486}
$

In accepting the C.M.V. constraint, we not only set the
dimensional projection equal to zero, but also eliminate the
R.C. variance in the dimensional direction from our uncertainty
estimates.
The projection of $\SiuuR$
onto the dimensionless subspace, $\Siuu_{dl}$, satisfies 
$\Siuu_{dl}= \SiuuR - \SiuuR \gh \gh^t \SiuuR /(\gh^t \SiuuR  \gh)$.

To evaluate the statistical uncertainty in the predicted
energy confinement for a given set of parameters,
we transform the tokamak's parameters
to the centered logarithmic variables, $\vec{x}^t$,
and take the interproduct with the covariance matrix of Table 4.
The centered $\xv^t$ variable is

$$
\left(
\left( \ln {R \over a} - 1.336 \right) \ ,
( \ln R - .528) \ ,
( \ln \kappa - .335) \ ,
1 \ , \right.
$$
$$
\left. ( \ln I_p  + .568) \ , ( \ln B_t - .573) \ ,
( \ln \nb - 1.501) \ ,
( \ln P - 1.071) \right) \ .
$$
The fourth index corresponds to the absolute constant in the
scaling law.

For I.T.E.R., we assume the following parameter value:
$M = 2.5$, $a = 2.15 m$, $R = 6.0m$, $\kappa = 2.2$, $I_p = 22 MA$,
$B_t = 4.85T$, $\nb = 14.0 \times 10^{19}$, $P_{\rm tot} = 160 MW$.
The resulting predicted confinement times is
4.65 sec with a statistical uncertainty
factor of 32\%.

For B.P.X., we use the following parameter values: $M = 2.5$,
$a = .8 m$, $R = 2.59m$, $\kappa = 2.2$, $I_p = 11.8MA$,
$B_t = 9.0T$, $\nb = 40 \times 10^{19}$, $P_{\rm tot} = 80 MW$.
We predict a B.P.X. H mode confinement time of 1.22 sec with an uncertainty
factor of 26 \%.

We note that the C.M.V. constraint reduces the estimated I.T.E.R. uncertainty
significantly more than the estimated B.P.X. uncertainty.
In the L mode database, size and magnetic field are strongly correlated,
i.e. the larger experiments have large magnetic fields,
especially TFTR and JT-60. JET has only a slightly larger magnetic field,
and therefore differs mostly in size. Since the H mode database has only
one large tokamak, the variance of the size exponent is crudely a factor
of three larger than the L mode size exponent. 

Our analysis of both the H and L mode databases underestimates the 
covariance of $B_t$ exponent by setting the number of degrees of freedom
equal to the number of tokamaks and not the number of tokamaks with
$B_t$ variation. However, the shrinkage factor is worse for the H
mode database (2/5) than for the L mode database (6/10).
Thus our random coefficient analysis finds that the size scaling
is more uncertain than the magnetic field scaling.

This explains why we find that unconstrained B.P.X. has a much smaller   
unconstrained uncertainty than I.T.E.R.. The collisional Maxwell Vlasov
constraint essentially couples the size and magnetic field scalings
and therefore reduces the I.T.E.R. uncertainty more than the B.P.X.
uncertainty.

\ns
\noindent
{\bf V. DISCUSSION}

Global scaling expressions, in particular, the ITER-89P  scaling$^6$,
have been successful in predicting the energy confinement in the
new series of experiments. The dimensionally constrained R.C. scaling
of Eq. 3 resembles the ITER-89P scaling in all parametric dependencies
except that our H mode scaling has a slightly stronger size scaling.

The random coefficient model is applicable when the
tokamak to tokamak differences are due to many small factors.
If, however, this tokamak to tokamak variation is attributable to one or more
important factors such as wall material or distance to the divertor plate,
statistics is of little help in analyzing confinement.


We find a predicted I.T.E.R. confinement time of 4.65 sec with a 
statistical uncertainty of $\pm 32 \%$
and a predicted B.P.X. confinement time of 1.22 sec 
with a statistical uncertainty of $\pm 26 \%$.
The unaccounted for uncertainties are discussed in Ref. [3].
When the constraint of collisional Maxwell Vlasov similarity is imposed,
the I.T.E.R. uncertainty is reduced from $34.9 \%$ to $ 32.4 \% $
while the B.P.X. uncertainty is slightly reduced from $ 27.9\%$ to $ 26.3\% $.

We find that the H mode data has a much larger intrinsically dimensional
component of the scaling than the corresponding L mode data.
This may be caused by the presence of other hidden variables.

We close on an optimistic note. We have repeated our constrained R.C.
analysis, however in the second stage regression, we have weighted the
larger tokamaks, D3D and JET even more heavily than the factor of two
used in the present analysis. We find that the large 
tokamak weighted scaling has an even stronger size scaling and
higher predicted confinement times for both I.T.E.R. and B.P.X.
than Eq. 3. This may indicate favorable departures of the confinement
time scaling from Eq. 3 for reactor size devices.

{\it Acknowledgment}

The author thanks Geoff Cordey and the H mode database group for
compiling the H mode database.
The author thanks C. Bolton, J. DeBoo, R. Goldston, 
O.J.W.F. Kardaun, S.M. Kaye,
and D. Post
for many useful discussions. Curt Bolton suggested the use of 
$P_{ohm}^*$ .

This work was supported under U.S. Department of Energy Grant No.
DE-FG02-86ER53223.

\np
\begin{center}
{\bf REFERENCES}
\end{center}

\begin{enumerate}
\item J.P. Christiansen, J.G. Cordey, K. Thomsen, et. al.,
``A Global Energy Confinement H-mode Database for I.T.E.R.",
to be submitted to {\it Nucl. Fusion}.
\item  Riedel, K.S.,
Nuclear Fusion, {\bf 30}, No. 4, p. 755, (1990).
\item Riedel, K.S., Kaye S.M., { Nuclear Fusion},
{\bf 30}, No. 4, p. 731, (1990).
\item Riedel, K.S., `` On dimensionally correct power law scaling
expressions for L mode confinement,"
accepted in { Nuclear Fusion},
{\bf 31}, No. ?, p. ?, (1991).
\item Goldston, R. J., G. Bateman, M.G. Bell, et. al., 
``Performance Projections for B.P.X.",
Bull. of Am. Phys. Soc., {\bf 35}, No. 9, p. 1920, poster 1P1, Oct. (1990).
\item Yushmanov, P., Takizuka, T., Riedel, K.S., Karadun, O., Cordey,
J., Kaye, S. and Post, D., Nucl. Fusion, {\bf 30}, p.1999, (1990)
\item Kadomtsev, B.B., Sov. J. Plasma Phys., Vol. 1, (1975) p.295
\item  Christiansen, J.P., Cordey, J.G., Kardaun, O.J.W.F.,
and Thomsen, K.,
to be submitted to Nucl. Fusion.

\end{enumerate}
\end{document}